\documentclass{PoSi}
\pdfoutput=1
\usepackage{lineno}
\usepackage{xspace}
\usepackage{subfigure}
\usepackage{amsmath,bm}
\usepackage{nicefrac}
\usepackage{url}
\usepackage{wrapfig}

\newcommand{\eV}{\ensuremath{\mbox{e\kern-0.1em V}}\xspace}
\newcommand{\GeV}{\ensuremath{\mbox{Ge\kern-0.1em V}}\xspace}
\newcommand{\MeV}{\ensuremath{\mbox{Me\kern-0.1em V}}\xspace}
\newcommand{\GeVc}{\ensuremath{\mbox{Ge\kern-0.1em V}\!/\!c}\xspace}
\newcommand{\GeVcc}{\ensuremath{\mbox{Ge\kern-0.1em V}\!/\!c^2}\xspace}
\newcommand{\AGeV}{\ensuremath{A\,\mbox{Ge\kern-0.1em V}}\xspace}
\newcommand{\AGeVc}{\ensuremath{A\,\mbox{Ge\kern-0.1em V}\!/\!c}\xspace}
\newcommand{\MeVc}{\ensuremath{\mbox{Me\kern-0.1em V}/c}\xspace}

\newcommand{\dedx}{\ensuremath{{\rm d}E\!/\!{\rm d}x}\xspace}
\newcommand{\pT}{\ensuremath{p_{\rm T}}\xspace}

\newcommand{\DPMJet}{{\scshape DPMJet}\xspace}
\newcommand{\Epos}{{\scshape Epos}\xspace}
\newcommand{\EposLong}{{\scshape Epos1.99}\xspace}
\newcommand{\QGSJetLong}{{\scshape QGSJetII-04}\xspace}
\newcommand{\DPMJetLong}{{\scshape DPMJet3.06}\xspace}
\newcommand{\SibyllLong}{{\scshape Sibyll2.1}\xspace}
\newcommand{\EposLHCLong}{{\scshape EposLHC}\xspace}

\newcommand{\NASixtyOne}{NA61\slash SHINE\xspace}
\newcommand{\CernVM}{\textsc{Cern\-\kern-0.05emVM}\xspace}

\title{Results from Pion-Carbon Interactions Measured by \NASixtyOne
  for Improved Understanding of Extensive Air Showers}

\ShortTitle{Pion-Carbon Interactions Measured by \NASixtyOne}

\author{\speaker{Alexander E.~Herv\'{e}} for the \NASixtyOne
  Collaboration\thanks{http://shine.web.cern.ch/content/author-list}\\ IKP,
  Karlsruhe Institute of Technology (KIT), Postfach 3640, D-76021
  Karlsruhe, Germany\\ E-mail: \email{alexander.herve@kit.edu}}

\abstract{The interpretation of extensive air shower measurements,
  produced by ultra-high energy cosmic rays, relies on the correct
  modeling of the hadron-air interactions that occur during the
  shower development. The majority of hadronic particles are produced
  at equivalent beam energies below the TeV range. \NASixtyOne is a
  fixed target experiment using secondary beams produced at CERN at
  the SPS. Hadron-hadron interactions have been recorded at beam
  momenta between 13 and 350\,\GeVc with a wide-acceptance
  spectrometer. In this contribution we present measurements of the
  spectra of charged pions and the $\rho^0$ production in
  pion-carbon interactions, which are essential for modeling of air
  showers.}

\FullConference{The 34th International Cosmic Ray Conference\\
                30 July -- 6 August, 2015\\
                The Hague, The Netherlands}

\begin{document}

\section{Introduction}
When cosmic rays collide with the nuclei of the atmosphere, they
initiate extensive air showers (EAS). The interpretation of EAS data
relies to a large extent on the understanding of these air showers,
specifically on the correct modeling of hadron-air interactions that
occur during the shower development. Experiments such as the Pierre
Auger Observatory~\cite{Abraham:2004dt},
KASCADE-Grande~\cite{Navarra:2004hi}, IceTop~\cite{IceTop} or the Telescope
Array~\cite{AbuZayyad:2012kk} use models for the interpretation of
measurements.  However, there is mounting evidence that current models
give a poor description of muon production in air showers (see
Refs. ~\cite{Arteaga-Velazquez:2013ira,muon_inclined,mpd2014}).

Unfortunately, there exist no comprehensive and precise particle
production measurements for the most numerous projectile in air
showers, the $\pi$-meson. Therefore, new data with pion beams at 158
and 350\,\GeVc on a thin carbon target (as a proxy for nitrogen) were
collected by the \mbox{\NASixtyOne} experiment at the CERN SPS.

Preliminary spectra of unidentified hadrons have been previously
derived from this data set and the spectra revealed discrepancies between the data
and predictions from generators for hadronic
interactions~\cite{ISVHECRI12_MU,Zambelli2013,Popov2014}.

In this contribution we will present the measurement of spectra of 
identified charged pions and $\rho^0$ mesons in $\pi^-$+C
interactions at 158 and 350\,\GeVc.

\section{The \NASixtyOne Experiment}
\NASixtyOne\footnote{SHINE = SPS Heavy Ion and Neutrino
Experiment}~\cite{na61_new} is a multi-purpose fixed target experiment
to study hadron production in hadron-nucleus and nucleus-nucleus
collisions at the CERN Super Proton Synchrotron (SPS). Among its
physics goals are precise hadron production measurements for improving
calculations of the neutrino beam flux in the T2K and Fermilab neutrino oscillation
experiment~\cite{Abe:2011ks} as well as for more reliable simulations
of hadronic interactions in air showers. Moreover, p+p, p+Pb and
nucleus+nucleus collisions are measured to study the
properties of the onset of de-confinement and search for the
critical point of strongly interacting matter (see
e.g.\ Ref.~\cite{Gazdzicki:2010iv}).

The \NASixtyOne Collaboration uses large time-projection-chambers
(TPCs) inherited from the NA49 experiment~\cite{Afanasev:1999iu} to
measure the charge and momentum of particles. The momentum resolution,
$\sigma(1/p)=\sigma(p)/p^2$, is about $10^{-4}$\,(\GeVc)$^{-1}$ at full
magnetic field and the tracking efficiency is better than 95\%.
A set of scintillation and
Cherenkov counters as well as beam position detectors upstream of the
spectrometer provide timing reference, identification and position
measurements of the incoming beam particles.  Particle identification
is achieved by measuring the energy loss along the tracks in the TPCs
and by determining their velocity from the time of flight provided by
large scintillator walls placed downstream of the TPCs.  The
centrality of nucleus-nucleus collisions can be estimated using the
measurement of the energy of projectile spectators with a
calorimeter~\cite{Golubeva:2012zz} located behind the time of flight
detectors. For nucleon-nucleus collisions, the centrality is
determined by counting low momentum particles from the target (so
called `gray protons') with a small TPC around the target~\cite{Marton:2014gqa}.

Data taking with the \NASixtyOne experiment started in 2007.  After a
first run with proton on carbon at 31\,\GeVc, the data acquisition
system was upgraded during 2008 to increase the event recording rate
by a factor of ${\approx}10$~\cite{Laszlo:2015lza}. In the last years,
a wealth of data has been recorded by the experiment at beam momenta
ranging from 13 to 350\,\GeVc with various beam particles and
targets. In this paper we present results obtained from a special run
with negative pions as beam particles and carbon as the target. Since
pions are the most numerous particles in an air shower, this data will
help to improve the interpretation of air showers at ultra-high energies.

\section{Spectra of Charged Pions}

\begin{wrapfigure}{r}{0.55\textwidth}
\centering
\includegraphics[width=\linewidth]{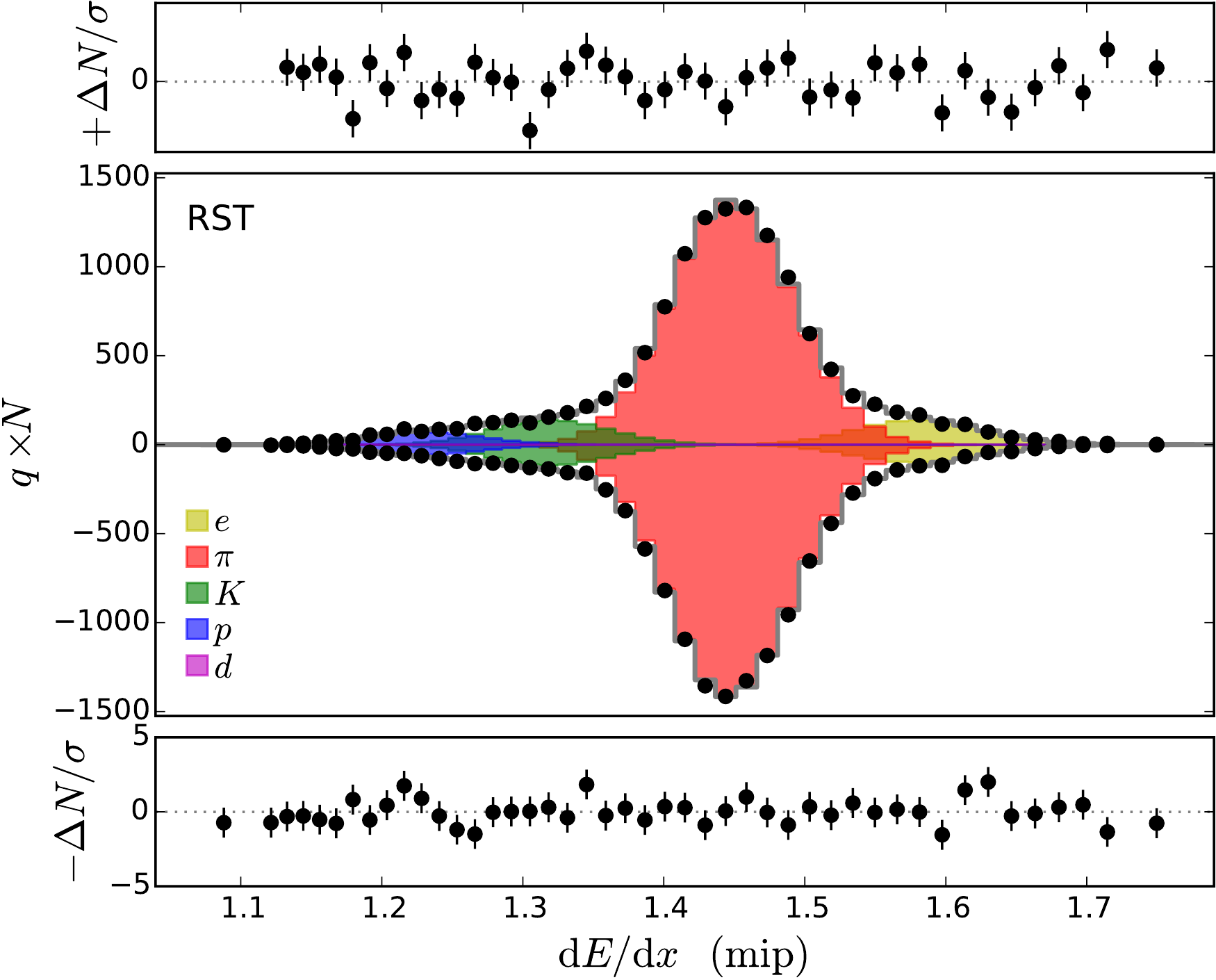}
\caption{Example of a \dedx fit
  ($\langle p \rangle = 28.4$\,\GeVc, $0<\pT<0.1$\,\GeVc). The middle
panel shows the energy deposit of positively and negatively charged
tracks. The fitted particles are indicated by colored histograms. Residuals
to the fit are shown in the upper and lower panels.}
\label{fig:dedx}
\end{wrapfigure}

For each track detected in the TPCs of NA61, the particle type can be
estimated by using the truncated mean of the energy that is deposited
per unit track length ($\dedx$) along the particle trajectory.  An
example of a $\dedx$-distribution in a specific bin in momentum $p$
and transverse momentum $\pT$ is shown in Fig.~\ref{fig:dedx}. As can
be seen, the distribution can be well described by the sum of the
energy loss distributions of electrons, protons, pions and kaons (see
Ref.~\cite{VerezPhD} for details) and given the fitted fraction of
each particle type, the corresponding number, $\Delta n$, of produced
tracks within each $p/\pT$-bin can be reconstructed.

This number is
then corrected for the detector acceptance, selection efficiency,
feed-down from weak decays and re-interactions in the target. The
latter two corrections are currently estimated using model predictions
(\EposLong, \QGSJetLong, \DPMJetLong) and they are typically well
below $5\%$, but can reach up to 20\% at low particle momenta. Overall,
the systematic uncertainty of the corrected number of tracks, $\Delta n'$,
is estimated to be ${\le}7\%$.

The average multiplicity of particles produced within a $p/\pT$-bin is
then obtained by dividing $\Delta n'$ by the total
number of events in which an interaction occurred, $N_\text{prod}$.
$N_\text{prod}$ is estimated by extrapolating the number of
recorded interaction triggers to full phase space. The correction
amounts to $(92.5\pm3.5)\%$ at 158\,\GeVc and $(92.5\pm4.0)\%$ at
350\,\GeVc, where the uncertainty was derived by running different generators
to evaluate the correction.

\begin{figure}[!t]
\centering
\includegraphics[width=0.9\linewidth]{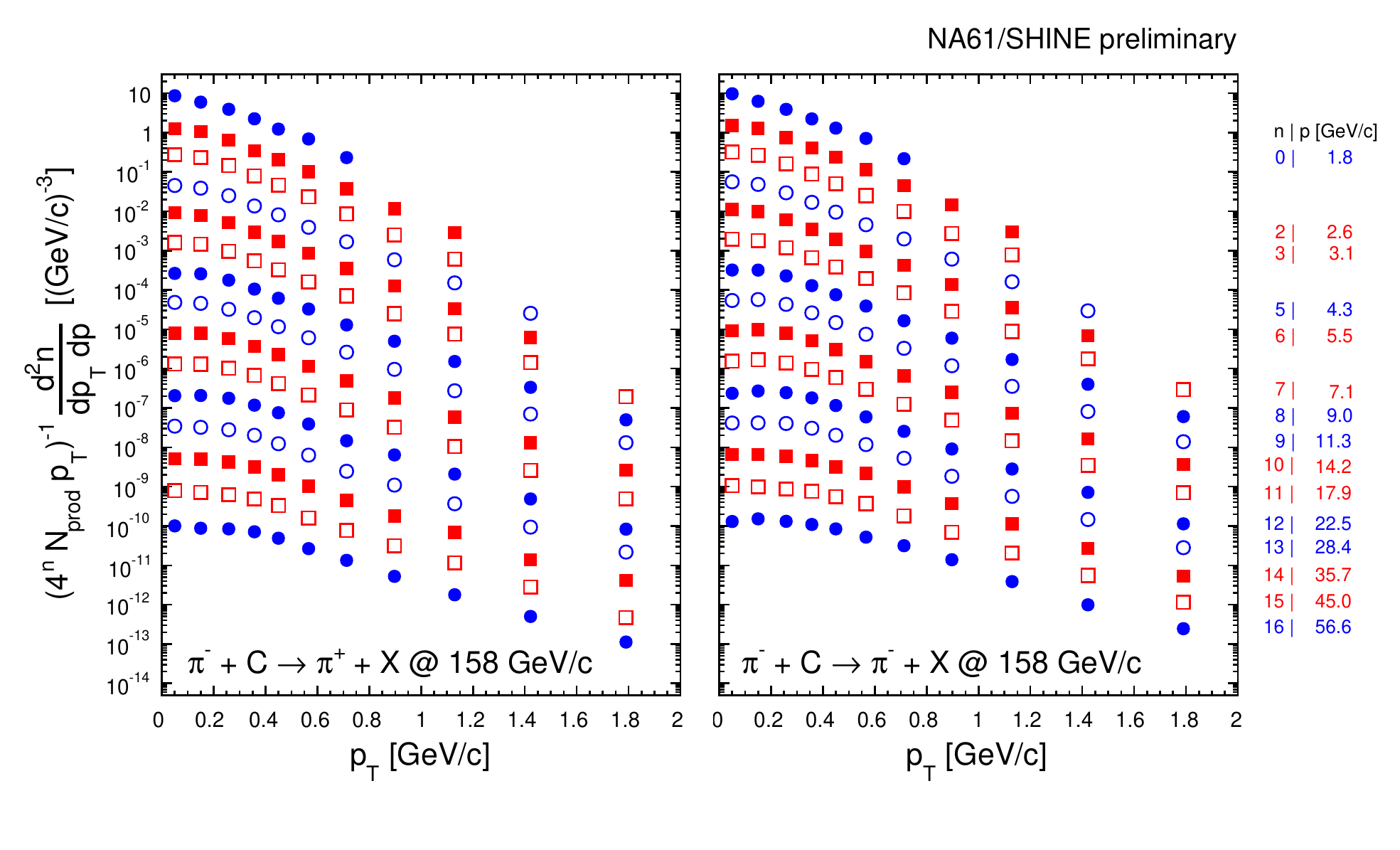}
\includegraphics[width=0.9\linewidth]{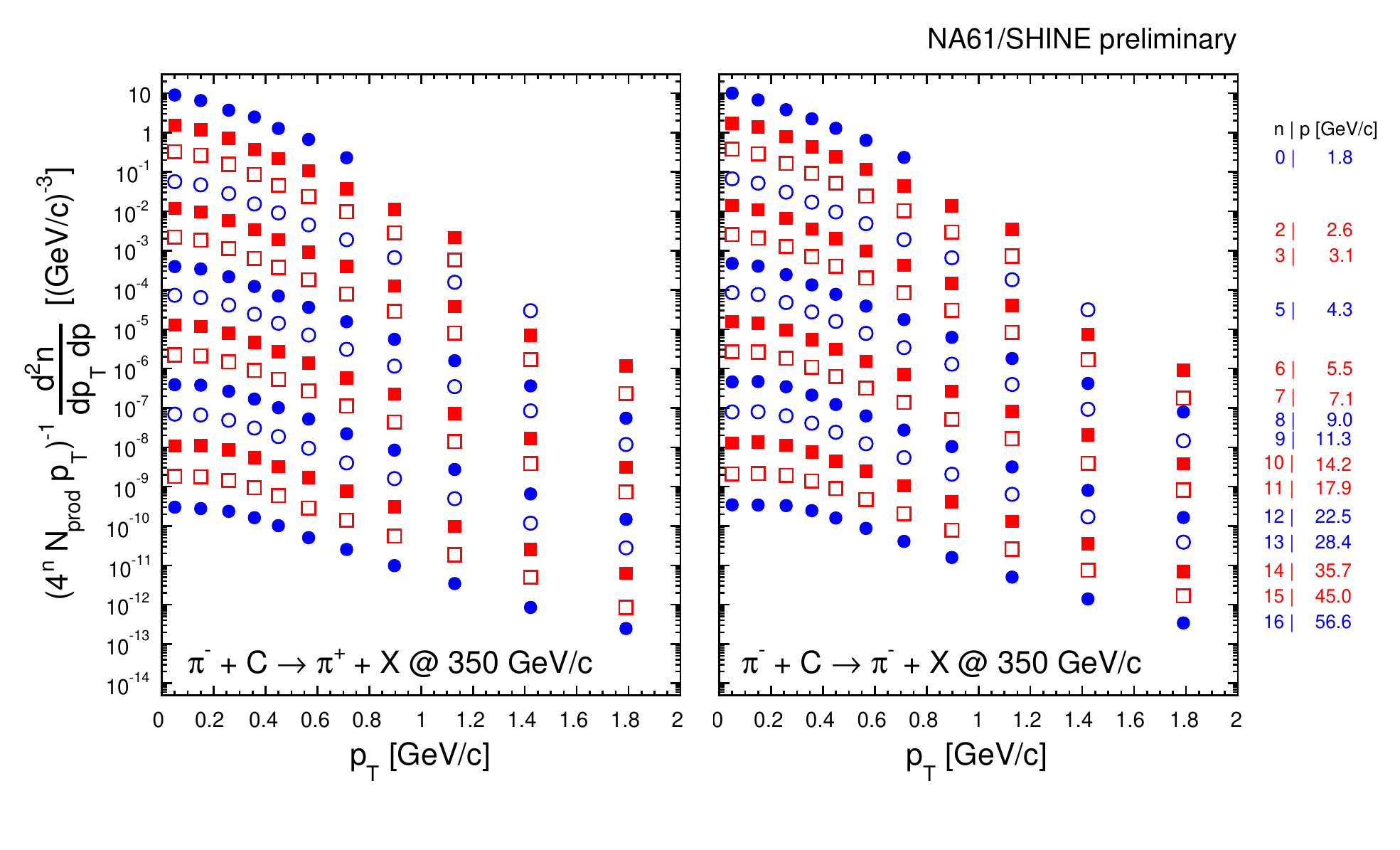}
\caption{Inclusive spectra of charged pions in $\pi^-$+C interactions
at beam energies of 158 and 350\,\GeVc. For better visibility, the spectra
from the $n$\textsuperscript{th}\xspace momentum bin are multiplied by a 
factor of $\nicefrac{1}{4^n}$. The momentum increases from top to bottom 
as indicated in the legend on the right. Systematic errors are less than $7\%$}
\label{fig:spectra}
\end{figure}

\begin{figure}[!p]
\centering
\includegraphics[width=0.85\linewidth]{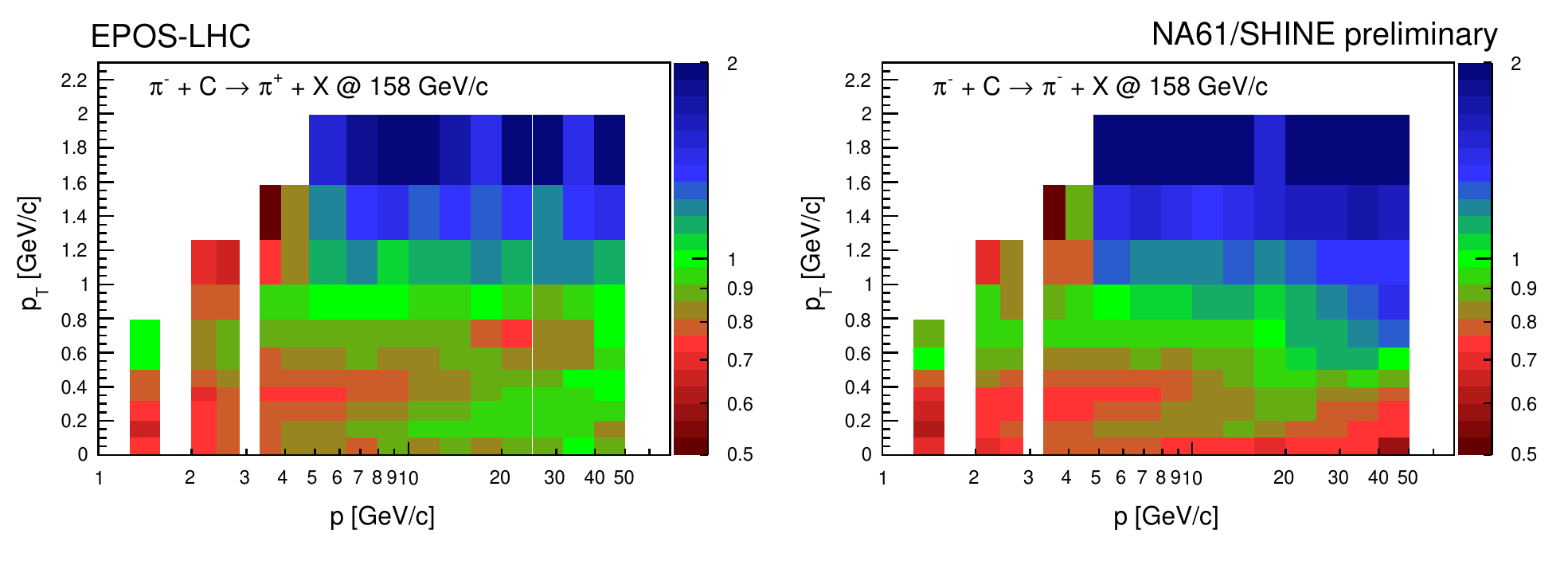}
\includegraphics[width=0.85\linewidth]{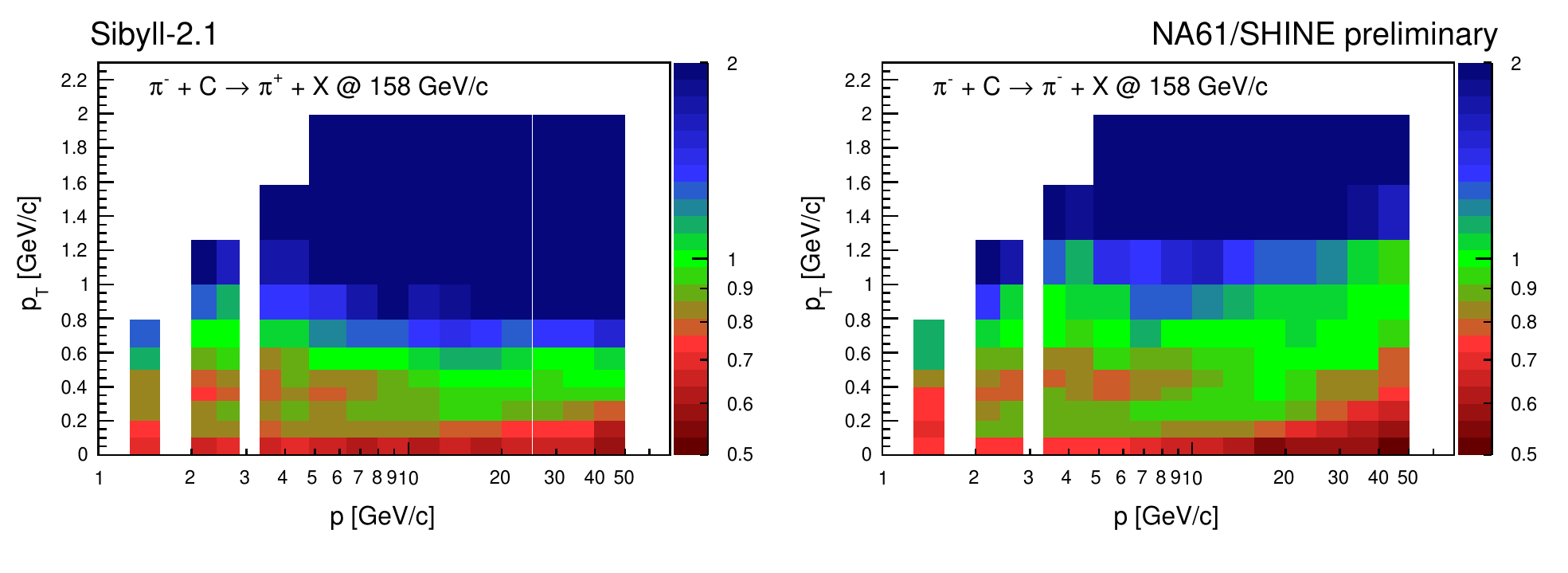}
\includegraphics[width=0.85\linewidth]{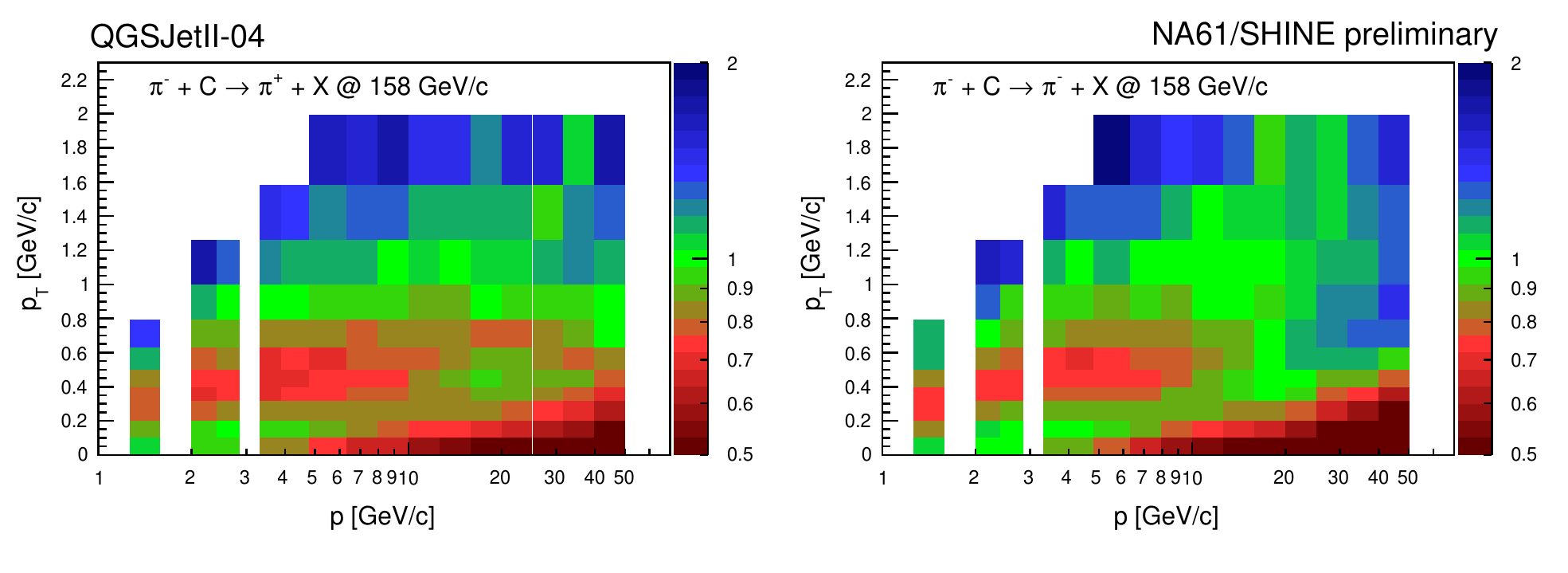}
\includegraphics[width=0.85\linewidth]{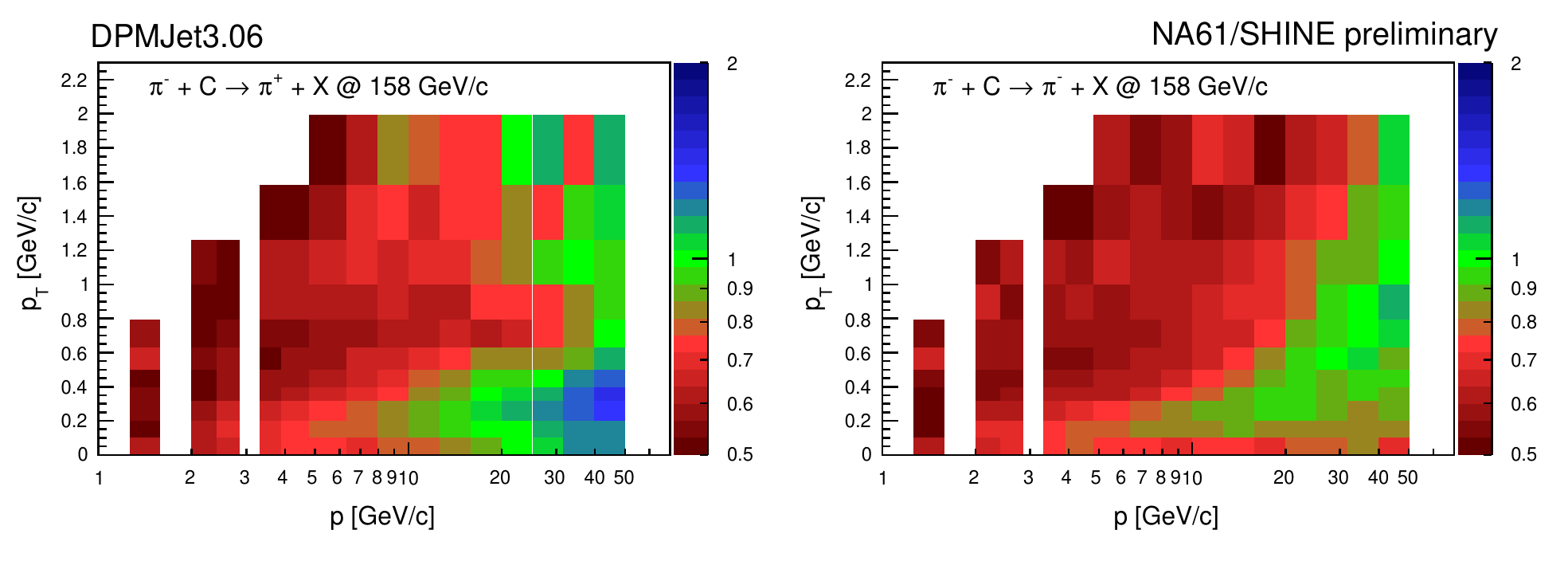}
\caption[model ratio]{Comparison of the measured production spectra of
  charged pions to predictions from hadronic interaction models used
  for the interpretation of cosmic ray data~\cite{eposlhc, sibyll2.1,
    qgsjetII, dpmjet}. The colors denote the ratio of data to
  simulation and the color scale is truncated at 0.5 and 2. The two
  empty $p$-bins at $p\lesssim 2$\,\GeVc and $p\gtrsim 3$\,\GeVc are
  momenta excluded from the analysis due to the ambiguity in 
  identification of pions.}
\label{fig:mcratio}
\end{figure}

The measured average multiplicities of charged pions are shown in
Fig.~\ref{fig:spectra} and the measurements are compared to
predictions of pion production in $\pi^-$+C at 158\,\GeVc from hadronic
interaction models in Fig.~\ref{fig:mcratio}. As can be seen, none of
the generators describes the data well.

\section[Spectra of $\rho^0$ Mesons]{Spectra of $\boldsymbol{\rho^0}$ Mesons}

The measurement of resonances in $\pi$+C is useful to constrain the
production of $\rho^0$ meson, which is important to predict the
number of muons observed in air showers as well as the baryon fraction (see
e.g.\ Ref.~\cite{Drescher:2007hc}).

In the inclusive $\pi^+\pi^-$ invariant mass spectra there is a large
combinatorial background, which dominates over the effective mass
distributions of individual resonances. The method used to estimate
the background is the so called charge mixing, which uses the
($\pi^+\pi^++\pi^-\pi^-$) mass spectra as an estimate of
the background.

The fitting procedure uses templates of the $\pi^+\pi^-$ mass
distribution for each resonance. These templates are constructed
by passing simulated $\pi$+C interactions, generated with the
\EposLong~\cite{epos} hadronic interaction model using \textsc{Crmc}~\cite{crmc} 
(v1.5.3), through the full NA61 detector Monte Carlo chain. All the 
cuts that are applied to the data are also applied to the templates. 
This method of using templates allows for the fitting of both resonances 
with dominant three body decays, such as the $\omega$, and resonances 
with non-$\pi^+\pi^-$ decays, such as the K$^{*0}$.
The data is split into bins of Feynman-$x$, $x_\text{F}$.

\begin{figure}[!t]
\centering \includegraphics[width=0.7\linewidth]{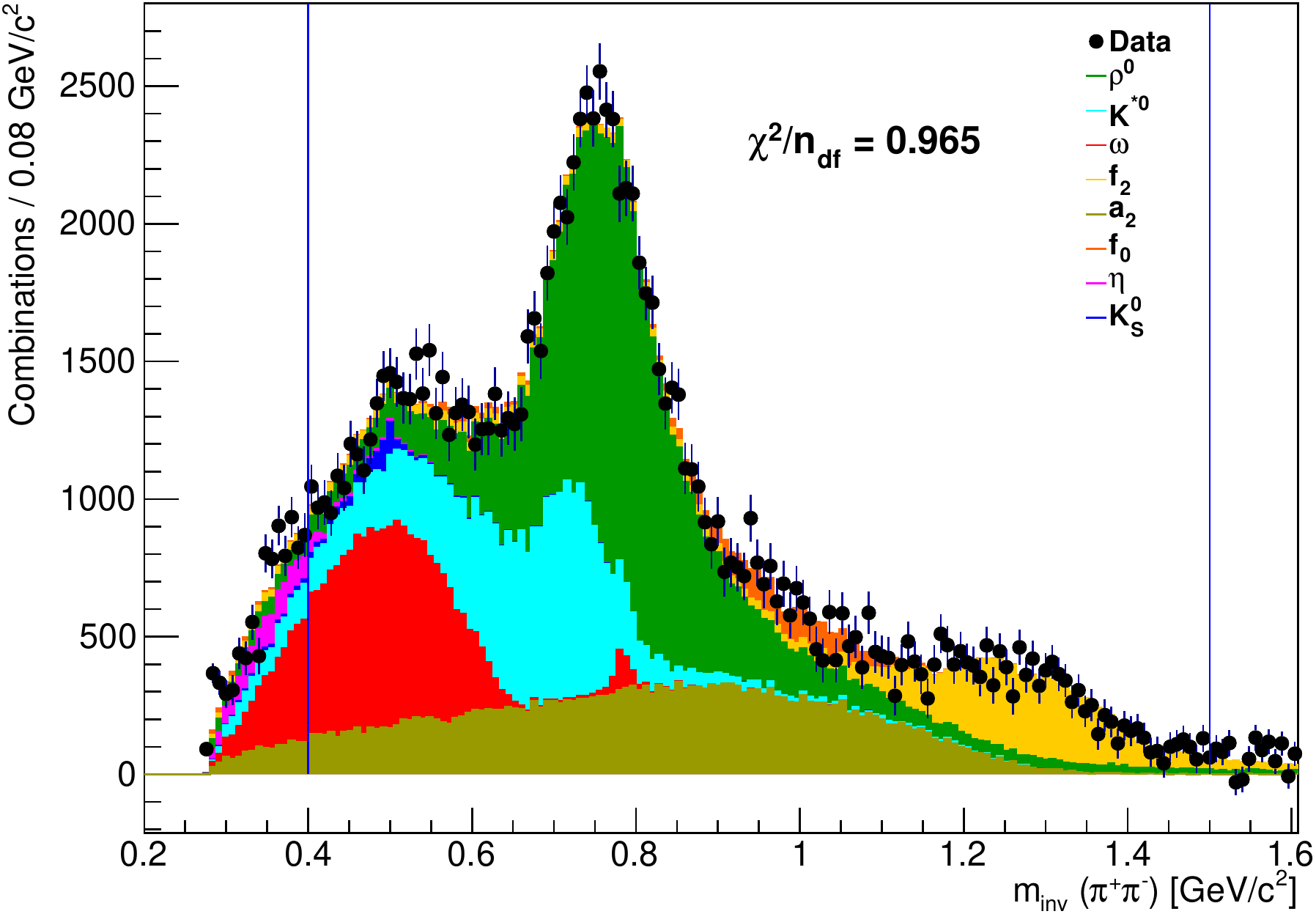}
\caption{$\pi^+\pi^-$ invariant mass distribution in $\pi^-$+C interactions
at 158\,\GeVc in the range $0.4 < x_\text{F} < 0.5$. Dots with error bars denote the
data and the fitted resonance templates are shown as filled histograms.
The vertical lines indicate the range of the fit.}
\label{fig:Example_Fit}
\end{figure}

The fit to the $\pi^+\pi^-$ mass spectrum is performed between
masses of 0.4\,\GeVc and 1.5\,\GeVc using the expression
\begin{equation*}
F(m) = \sum_i \beta_i \, T_i(m),
\end{equation*}
where $\beta_i$ is the relative contribution for each template, $T_i$, used.
An example of one of these fits can be seen in Fig.~\ref{fig:Example_Fit},
The templates in the fit are the background found from charge mixing and
the following resonances: $\rho^0$, K$^{*0}$, $\omega$, f$_2$, f$_0$ (980),
a$_2$, $\eta$ and K$^0_S$.

The fitting method is validated by applying the same procedure to
the simulated data set which was used to construct the templates
for the fit.
For the majority of $x_\text{F}$ bins there is good agreement between the fit
and the true value, with some discrepancies for larger $x_\text{F}$ bins of up to 20\%. This
bias is corrected for in the final analysis. The data is also corrected for
losses due to the acceptance of the detector, as well as any bias due to
the cuts used and any reconstruction efficiencies. Apart from the acceptance,
these corrections are typically less than 20\%.

\begin{figure}[!t]
\centering \includegraphics[width=0.8\linewidth]{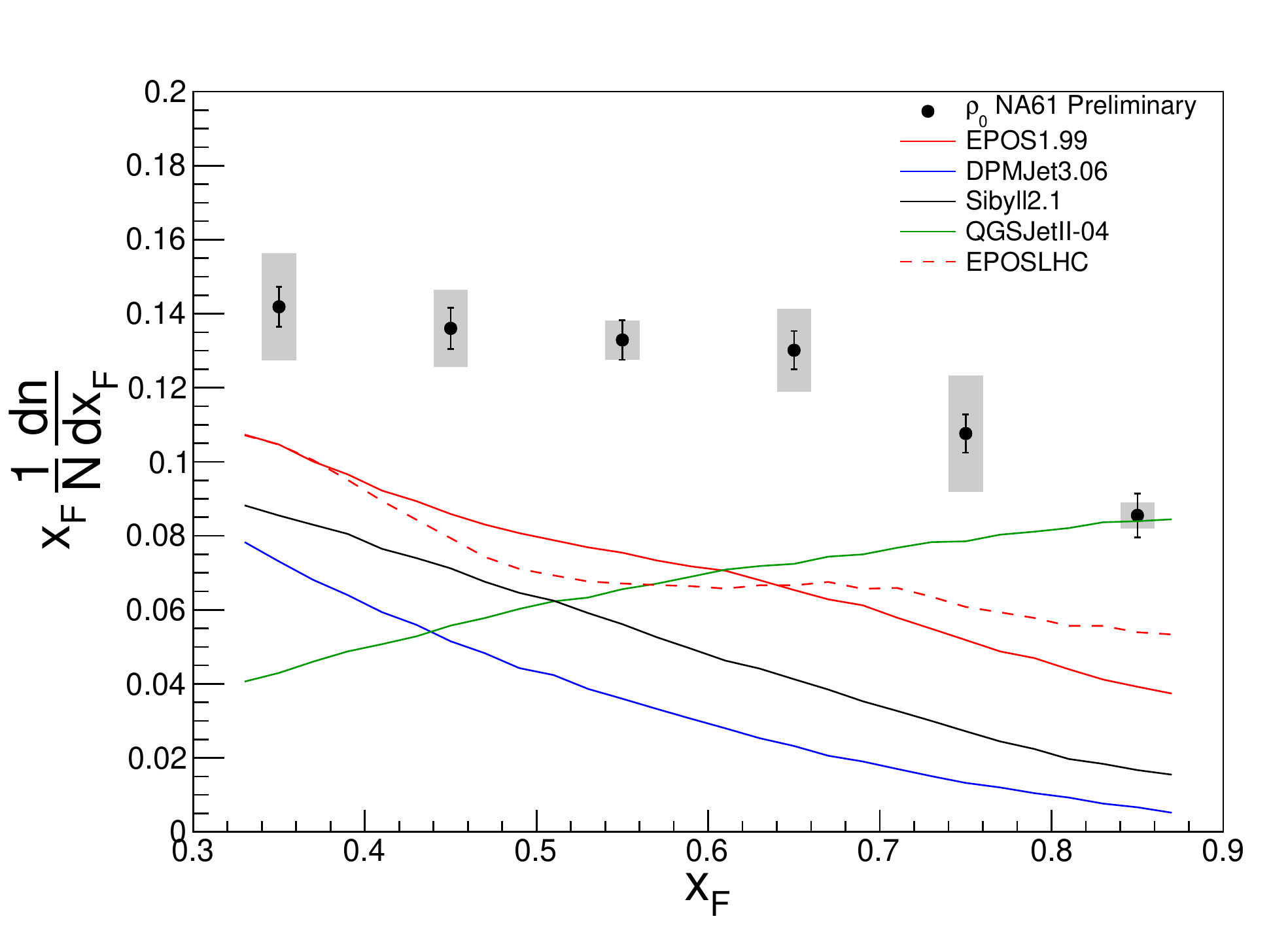}
\caption{Average multiplicity of the $\rho^0$ meson in $\pi$+C at
$p_\text{beam}=158\,\GeVc$ as a function of Feynman-$x$. The bars
show the statistical errors; the bands indicate systematic errors.
The lines depict predictions of hadronic interaction models: red -
\EposLong, blue - \DPMJetLong, black - \SibyllLong, green -
\QGSJetLong, dashed red - \EposLHCLong.}
\label{fig:Rho_Result}
\end{figure}

The average multiplicity of $\rho^0$ mesons is presented in
Fig.~\ref{fig:Rho_Result}. Also shown are predictions by
\EposLong~\cite{epos}, \DPMJetLong~\cite{dpmjet},
\SibyllLong~\cite{sibyll2.1}, \QGSJetLong~\cite{qgsjetII} and
\EposLHCLong~\cite{eposlhc}. It can be seen that there is an
underestimation of the $\rho^0$ yield for almost all hadronic interaction
models, with the exception of \QGSJetLong for $x_\text{F} > 0.8$.
It is interesting to note that while \QGSJetLong and \EposLHCLong were
tuned to NA22 $\pi^+$+p data \cite{NA22}, there is an underestimation in
$\pi^-$+C.

Systematic errors are estimated by comparing correction factors
for different hadronic interaction models (\Epos and \DPMJet), comparing
the correction for the bias using different background estimates and
varying the cuts applied to the data. The systematic is dominated by
the background estimates, up to 14\%, where as the other errors are
less than 4\%. Other sources of uncertainty, such as using templates
from a different model, are found to be much smaller.

\section{Conclusions and Outlook}
In this article, we summarized results from pion-carbon interactions
measured with the multi-purpose experiment \NASixtyOne at the CERN
SPS, which are of importance for the modeling of cosmic ray air showers.

The comparisons to hadronic interaction models shown in this article
suggest that these models require further tuning to reproduce the charged
pion and $\rho^0$ meson spectra.

It is planned to further refine both analyses presented here,
including the measurement of inclusive spectra of charged kaons and
protons as well as the study of the multiplicities of other resonances in
addition to the $\rho^0$.

\clearpage

{\footnotesize{{\bf Acknowledgment:} This work was supported by the
Hungarian Scientific Research Fund (grants OTKA 68506 and 71989),
the Janos Bolyai Research Scholarship of the Hungarian Academy of
Sciences, the Polish Ministry of Science and Higher Education
(grants 667/N-CERN/2010/0, NN 202 48 4339 and NN 202 23 1837), the
Polish National Center for Science (grant 2011/03/N/ST2/03691), the
Polish National Center for Science (grant 2013/11/N/ST2 /03879) the
Foundation for Polish Science -- MPD program, co-financed by the
European Union within the European Regional Development Fund, the
Federal Agency of Education of the Ministry of Education and Science
of the Russian Federation (grant RNP 2.2.2.2.1547), the Russian
Academy of Science and the Russian Foundation for Basic Research
(grants 08-02-00018 and 09-02-00664), the Ministry of Education,
Culture, Sports, Science and Technology, Japan, Grant-in-Aid for
Scientific Research (grants 18071005, 19034011, 19740162, 20740160
and 20039012), the German Research Foundation (grant GA 1480/2-2), 
Ministry of Education and Science of the Republic of Serbia (grant 
OI171002), Swiss Nationalfonds Foundation (grant 200020-117913/1) 
and ETH Research Grant TH-01 07-3.}

\end{document}